\documentclass[12pt]{article}

\usepackage{axodraw4j}
\usepackage{pstricks}
\usepackage{color}
\usepackage{latexsym,amsmath,amssymb}
\usepackage{graphicx}
\usepackage{slashed}  
\usepackage{bm}
\usepackage{epsfig}
\usepackage{dcolumn}
\usepackage{cite}

\usepackage{bm}
\usepackage{latexsym}
\usepackage{amsmath}
\numberwithin{equation}{section}

\usepackage{amsmath}
\numberwithin{equation}{section}
\def\ee{\end{equation}}
\def\ba{\begin{eqnarray}}
\def\ea{\end{eqnarray}}

\def\bq{\begin{quote}}
\def\eq{\end{quote}}

 at 10truept

\newcommand{\beq}{\begin{equation}}
\newcommand{\eeq}{\end{equation}}
\newcommand{\beqa}{\begin{eqnarray}}
\newcommand{\eeqa}{\end{eqnarray}}
\newcommand{\bea}{\begin{eqnarray}}
\newcommand{\eea}{\end{eqnarray}}
\newcommand{\p}{\partial}
\newcommand{\al}{\alpha}
 \newcommand{\be}{\beta}
 \newcommand{\ep}{\epsilon}
\newcommand{\e}{\epsilon}
\newcommand{\si}{\sigma}

\newcommand{\lmk}{\left(}
\newcommand{\rmk}{\right)}
\newcommand{\lkk}{\left[}
\newcommand{\rkk}{\right]}

\newcommand{\lb}{\left|}
\newcommand{\rb}{\right|}

\newcommand{\vect}[1]{\bm{\mathrm{{#1}}}}

\newcommand{\hf}{\frac{1}{2}}

\def\lesssim{~\mbox{\raisebox{-.6ex}{$\stackrel{<}{\sim}$}}~}

\def\ltap{\ \raise.3ex\hbox{$<$\kern-.75em\lower1ex\hbox{$\sim$}}\ }
\def\gtap{\ \raise.3ex\hbox{$>$\kern-.75em\lower1ex\hbox{$\sim$}}\ }
\def\gl{\ \raise.5ex\hbox{$>$}\kern-.8em\lower.5ex\hbox{$<$}\ }
\def\roughly#1{\raise.3ex\hbox{$#1$\kern-.75em\lower1ex\hbox{$\sim$}}}

\newcommand{\txt}{\textnormal}
\newcommand{\til}{\tilde}
\newcommand{\Fab}{F_{\al \be}}
\newcommand{\Fmn}{F_{\mu \nu}}
\newcommand{\FMN}{F^{\mu \nu}}

\newcommand{\tFMN}{\tilde{F}^{\mu \nu}}
\newcommand{\sg}{\sqrt{-g}}
\newcommand{\pr}{^{\prime}}
\newcommand{\ppr}{^{\prime \prime}}

\newcommand{\lam}{\lambda}

\newcommand{\nab}{\nabla}
\newcommand{\nam}{\nabla_\mu}
\newcommand{\Ra}{\Rightarrow}
\newcommand{\ra}{\rightarrow}

\newcommand{\ze}{\zeta}
\newcommand{\nn}{\nonumber}
\def\pa{\partial}
\def\k{{\vec k}}
\newcommand{\cR}{{\cal R}}

\newcommand{\s}{\sigma}
\newcommand{\pI}{\delta \phi_I}
\newcommand{\ddpI}{\ddot{\delta \phi_I}}
\newcommand{\dpI}{\dot{\delta \phi_I}}

\def\baq{\begin{eqnarray}}
\def\eaq{\end{eqnarray}}

\begin{document}
\thispagestyle{empty}
\begin{titlepage}
\nopagebreak

\title{  \begin{center}\bf Universal Constraints on Axions from Inflation \end{center} }

\vfill
\author{Ricardo Z. Ferreira\footnote{ferreira@cp3.dias.sdu.dk} ~ and~  Martin S. Sloth\footnote{sloth@cp3.dias.sdu.dk}
}
\date{ }

\maketitle

\begin{center}
\vspace{-0.7cm}
{\it  CP$^3$-Origins, Center for Cosmology and Particle Physics Phenomenology}\\
{\it  University of Southern Denmark, Campusvej 55, 5230 Odense M, Denmark}

\end{center}
\vfill
\begin{abstract}
We consider the presence of an axion like particle, $\sigma$, with a generic $CP$ violating axial coupling of the form $ (\alpha \, \sigma/f) F \tilde{F}$, where $F_{\mu\nu}$ is the gauge field strength of a generic abelian $U(1)$ gauge group, not necessarily associated with the standard electromagnetism, and $f$ is the decay constant of the axion. It has previously been demonstrated that if the axion is identified with the inflaton, such an interaction can lead to measurable cosmological signatures (non-Gaussian modifications of the curvature perturbation spectrum) depending on the parameter $\xi= \alpha \dot{\sigma}/ (f H)$. In the present paper we will show that the generation of curvature perturbation at horizon crossing due to the axial coupling has a universal form and remains unmodified in terms of the $\xi$ parameter even if the axion, $\sigma$, is not identified with the inflaton. As a consequence, it does not appear to be possible to generate CMB tensor perturbations through this mechanism, larger than the vacuum ones, without violating the observational constraints unless we combine this mechanism with a curvaton or if the $\sigma$ field becomes heavy and decays during inflation. Even in this last case there are non-trivial constraints coming from the slow-roll evolution of the curvature perturbation on super horizon scales which should be taken into account. We also comment on implications for inflationary models where axions play an important role as, for example, models of natural inflation where more than one axion are included and models where the curvaton is an axion.
 \end{abstract}
\noindent
DNRF90
\hfill \\
\vfill
\end{titlepage}

\section{Introduction}

It is believed that pseudo-scalar axion-like particles might play a crucial role for our understanding of inflation (see \cite{Freese:1990rb,Adams:1992bn,Kim:2004rp,Dimopoulos:2005ac,Silverstein:2008sg,Kaloper:2008fb,Ross:2009hg,Kaloper:2011jz,Pajer:2013fsa,Kehagias:2014wza} for some discussions of this point). In large field models of inflation a slightly broken continuous shift-symmetry protects the flatness of the inflaton potential from being spoiled by large loop corrections. The axion is identified with the Nambu-Goldstone boson of the spontaneously broken shift symmetry \cite{Freese:1990rb} . Topological effects also explicitly break the shift symmetry and generate a periodic potential for the axion. Axions will also couple to gauge fields, with a $CP$ violating coupling of the form \cite{Anber:2006xt, Anber:2009ua}
\beq
{\cal{L}}_\txt{int} =-\frac{\al \si}{4f}  F_{\mu \nu}\til{F}^{\mu \nu}~.
\label{Lint1}
\eeq
Notice this term in the Lagrangian is a total derivative if $\si$ is time-independent. We therefore expect all dynamical effects to be suppressed by time-derivatives of the axion. On the other hand, if the axion is slowly rolling one would expect non-trivial effects of the coupling in Eq. (\ref{Lint1}).

In fact, it is well known that if the axion, $\sigma$, is identified with the slowly rolling inflaton, the axial coupling above can have important observational implications due to the generation of non-Gaussianity \cite{Barnaby:2010vf} and the enhancement of the gauge field \cite{Anber:2006xt}. Writing the axion field in terms of an homogenous background part and a perturbation, $\sigma(t,\vec x) =\sigma(t) +\delta\sigma(t,\vec x)$, and identifying $\sigma$ with the inflaton, there will be a direct coupling of the inflaton perturbation and the gauge field of the form \footnote{Constraints on gauge field production from inflation with a scalar instead of a pseudo-scalar coupled to gauge fields through the interaction $\lambda(\si) F_{\mu \nu} F^{\mu \nu}$ was studied thoroughly  in \cite{Nurmi:2013gpa}.}  
\beq
{\cal{L}}_{\delta\si AA} =-\frac{\al \delta\si}{4f}  F_{\mu \nu}\til{F}^{\mu \nu}~.
\label{Lint2}
\eeq
Therefore there will be one-loop effects, with a gauge field running in the loops, which are enhanced by the resonant production of the gauge field parametrized by $\xi= \al \dot{\si}/ (f H)$, and should be distinguished from the resonant non-Gaussianity also present in many models of axion inflation due to superimposed fast oscillatory contributions to the axion potential itself \cite{Hannestad:2009yx,Flauger:2010ja}.

Interestingly, it has been demonstrated that similar loop effect also leads to an enhancement of tensor modes \cite{Sorbo:2011rz}. It is intuitively simple to understand that this effect is smaller. Tensor modes are only gravitationally coupled to the gauge field, and, therefore, the contribution from the enhanced gauge field in the loops to tensor perturbations is suppressed compared to the contribution to inflaton perturbations which couple directly to the gauge field, if the inflaton is identified with the axion. In this type of models large contributions to tensor correlation functions are, thus, always associated with large non-Gaussian contributions to the inflaton correlation functions. 

On the other hand, we might easily expect many axions to be around during inflation \cite{Arvanitaki:2009fg}, and not all of them need to contribute to the background evolution. In the case where the axion is not identified with the inflaton, but instead considered to be some isocurvature field, then, given that the direct coupling of the inflaton perturbation with the gauge fields in Eq. (\ref{Lint2}) is absent, it seems to be possible to have a large enhancement of tensor correlation functions without an associated large non-Gaussianity in the inflaton correlation functions. This is an interesting idea, since this could imply that a large B-mode in the CMB might not necessarily be related to the vacuum contribution to the tensor modes, but instead be sourced by this mechanism \cite{Mukohyama:2014gba}\footnote{Other similar ideas has been proposed in \cite{Senatore:2011sp,Biagetti:2013kwa}.}. If this is the case, a measurement of primordial B-modes would not be enough to determine the scale of inflation. Since the tensor modes would be very non-Gaussian, it was also proposed as a way of generating large tensor non-Gaussianities without scalar non-Gaussianities \cite{Cook:2013xea}.
In this light we revisit the cosmological signatures of axions present during inflation. We will show that even if the inflaton is only gravitationally coupled to the gauge fields, the presence of a sub-dominant axion coupled to the gauge field will still introduce a significant coupling between the gauge field and the curvature perturbation which can only be erased if the field decays quickly after the first observable modes become super horizon. This is because the fluctuations of the inflation and the axion are not gauge invariant and the physical gauge invariant curvature and entropy perturbations are, in general, linear combinations of the inflaton and axion field fluctuations. Only in one gauge the entropy perturbation can be directly associated with the axion field fluctuation, and we will show that even in this gauge the curvature perturbation couples to the gauge fields in a universal fashion in terms of the parameter $\xi$.

One way to see the universal form of the coupling between the curvature perturbation and the gauge field is to make the following observation. The axial coupling is a total derivative unless $\dot\si\neq 0$. In this case the equations of motion of the gauge field are conformal invariant, and to leading order one might therefore expect the gauge field only to be sensitive to variations in the background through axial coupling, which we can expand in variations of the background as \cite{Jain:2012ga}
\beq\label{gchange}
\sigma =\sigma |_{\delta\ln a=0} +\frac{ d\sigma}{d\ln a}\Big|_{\delta\ln a=0}\delta\ln a+\dots = \sigma|_{\zeta=0} +\frac{ d\sigma}{d\ln a}\Big|_{\zeta=0}\zeta +\dots
\eeq
Using $d\sigma/d\ln a = \dot\si/H$, we have in the comoving gauge 
\beq
{\cal{L}}_{\zeta AA} =\frac{\xi}{4}  \zeta F_{\mu \nu}\til{F}^{\mu \nu}~.
\label{Lint0}
\eeq
This can also be understood more formally as a gauge transformation from the spatially flat gauge to the comoving gauge. In the spatially flat gauge the spatial curvature fluctuation vanishes ($\zeta =0$) and the adiabatic scalar fluctuation is instead given by the inflaton perturbation $\phi(t,\vec x) = \phi_0(t) +\delta\phi(t,\vec x)$, while the perturbation of the axion field is $\sigma(t,\vec{x}) = \sigma_0(t) +\delta\sigma(t,\vec x)$. One can make then a transformation into the comoving gauge with $\delta\phi=0$ by employing the invariance under time reparameterizations. Under a time translation given by the vector $\xi_\mu = (T,0,0,0)$, the field perturbations transform as \cite{Bruni:1996im,Jarnhus:2007ia}
\beq
\delta\phi(x_\mu+\xi_\mu)= \delta\phi(x_\mu)+ \sum_{n=1}^\infty\frac{(\xi_\mu\p^\mu)^n}{n!}\phi(x_\mu)~,\qquad \delta\si(x_\mu+\xi_\mu)= \delta\si(x_\mu)+ \sum_{n=1}^\infty\frac{(\xi_\mu\p^\mu)^n}{n!}\si(x_\mu)~.
\eeq
To linear order the choice $T= -\delta\phi/\dot\phi$ will imply $\delta\phi(x_\mu+\xi_\mu)=0$, while solving for the spatial curvature perturbation gives $\zeta = HT$, which implies that in the new gauge, we can write 
\beq
\delta\si(x_\mu+\xi_\mu)= \delta\si(x_\mu) +\frac{\dot\si}{H}\zeta(x_\mu) +\dots
\eeq
which explains immediately how (\ref{Lint0}) is obtained from (\ref{Lint2}), and that when we go to the comoving gauge we are making a space-time dependent change of the time variable, such that $\sigma$ is shifted like in Eq. (\ref{gchange}). 

It is important to point out that only in the comoving gauge can we identify the curvature perturbation, $\zeta$, with the gauge invariant adiabatic curvature perturbation and the axion perturbation $\delta\si$ with the gauge invariant entropy peturbation. Below we will verify Eq. (\ref{Lint0}) in more details by deriving it directly in both the comoving and the flat gauge, and show explicitly that the leading interaction between the curvature perturbation and the gauge field has this form, independently of whether we identify the axion with the inflaton or not, as the above intuitive argument is indicating.

This paper is organized as follows. In section \ref{review} we review the mechanism by which the gauge fields are resonantly enhanced due to the coupling with a pseudo-scalar while in section \ref{est} an estimate of the contribution to cosmological correlators is given. In section \ref{inter} we compute the relevant interaction terms at cubic order in two different gauges and in section \ref{SHevSec} we study the superhorizon evolution of the curvature perturbation. Finally, in section \ref{impli} we discuss the cosmological implications of the universal coupling to scalar curvature perturbations and we conclude in section \ref{conc}.

\section{Gauge field production} \label{review}

In this section we review the mechanism by which gauge fields are resonantly enhanced. The existence of a pseudo-scalar, $\si$, during inflation allows for the presence of an axial coupling with strength $\al/f$ with a $U(1)$ gauge field, where $\al$ is related to the coupling of the axion to the gauge field and $f$ is the decay constant. Therefore, the following two terms in the Lagrangian are, possibly, of importance
\footnote{We choose natural unities where $M_{p}=1$.}
\baq
{\cal{L}}_\txt{int} =-\frac{\al \si}{4f}  F_{\mu \nu}\til{F}^{\mu \nu} \quad \txt{and} \quad {\cal{L}}_\txt{kin} =-\frac{1}{4} \FMN \Fmn,
\label{Lint}
\eaq
where $\Fmn = \pa_{\mu} A_{\nu}-\pa_{\nu} A_{\mu}$ is the EM field strength\footnote{The gauge field $A_{\mu}$ can represent the EM field or a dark $U(1)$ gauge field, although for simplicity we refer to it as EM.} and $\tFMN = \hf \frac{\epsilon^{\mu \nu \al \be}}{\sg} \Fab$ is its dual with $\epsilon^{0123}=1$. Notice that $\si$ could be the inflaton or any other pseudo-scalar.

For a FLRW space-time with metric 
\beq
ds^2=a^2(\tau) \lmk -d\tau^2+d\vec{r}^{\,2} \rmk,
\eeq
the background inflationary dynamics satisfies the equation
\beq
3 {\cal{H}}^2= \hf \phi^{\prime 2}-\hf (\nabla \phi)^2+a^2 V(\phi)+ \frac{a^2}{2} \left( \vec{E}^2 +\vec{B}^2 \right), \quad '\equiv \frac{\partial}{\partial \tau}
\eeq
where ${\cal{H}}\equiv a\pr/a$ and we denoted $\phi$ as the inflaton. We have also defined the electric and magnetic fields in the standard way as, respectively, $\vec{E}=-a^{-2} \vec{A}'$ and $\vec{B}= a^{-2} \nabla \times \vec{A}$. The equation of motion for the gauge field $A_\mu$ in the presence of the axial coupling is given by
\baq
\nabla_\mu F^{\mu \nu} + \nabla_\mu \lmk \frac{\al \si}{f} \til{F}^{\mu \nu} \rmk=0.
\label{Aeom0}
\eaq
If we choose the Coulomb gauge, where $A_0=\vec{\nabla} \cdot \vec{A}=0$, and quantize $A_\mu$ in the basis of the circular polarization vectors\footnote{The circular polarization vectors $ \vec{e}_\pm$ obey  $\k \cdot \vec{e}_{\pm} (\k)=0, \k \times \vec{e}_{\pm} (\k)=\mp ik \vec{e}_{\pm} (\k), \vec{e}_{\pm} (-\k)=\vec{e}_{\pm}(\k)^*$ and are normalized to $\vec{e}_{\lam}(\k)^* \cdot \vec{e}_{\lam \pr}(\k)=\delta_{\lam, \lam \pr}$.}, the equation of motion becomes
\beq
A_{\pm}(\tau,k) \ppr + \lmk k^2 \pm \frac{2k\xi}{\tau}\rmk A_{\pm}(\tau,k)=0,\quad \xi \equiv \frac{\al \dot{\si}}{2fH}=\frac{\al}{f} \sqrt{\frac{\e_\si}{2}}
\label{eq:A_finaleom}
\eeq
where $\e_\si \equiv \dot{\si}^2/(2H^2)$. If $\si$ is the inflaton then $\e_\si=\e$ is the first slow roll parameter. We will assume that the field $\si$ rolls slowly and therefore we can neglect the time dependence of $\xi$. Therefore, the results are accurate up to time variations of $\xi$.
Assuming $\dot{\si}>0$ and  $\tau<0$ during inflation, there is an enhancement of $A_+(\tau,k)$ whenever $ -k\tau<2\xi$ \footnote{Given that the $A_-$ component does not get excited we will neglect this component and, for simplicity, refer to $A$ as the (+) component.}.
In \cite{Anber:2009ua}, an accurate solution for Eq. (\ref{eq:A_finaleom}) in the interval $(8\xi)^{-1}\lesssim -k\tau \lesssim 2\xi$ was found to be
\beq
A(\tau,k)\simeq \lmk \frac{-\tau}{2^3k\xi} \rmk^{1/4} e^{\pi \xi-2\sqrt{-2\xi k \tau}} \quad \txt{and} \quad A'(\tau,k) \simeq \lmk \frac{-k \xi}{2 \tau} \rmk^{1/4} e^{\pi \xi -2 \sqrt{-2 \xi k \tau}}.
\label{approxA}
\eeq
This time interval is of importance because it corresponds to the interval where the gauge modes are effectively excited. After that time, when $-k\tau \ra 0 $, the gauge field saturates to a constant value and therefore the EM energy density decreases with the scale factor adiabatically, as $a(\tau)^{-4}$. Basically, the field gets an enhancement around the time of horizon crossing.

This picture is consistent as long as the inflationary dynamics remains unchanged. Therefore the energy stored in the gauge fields, which is given by \cite{Anber:2009ua} \footnote{As discussed in \cite{Anber:2009ua, Barnaby:2011vw} the modes are in their vacuum state for $-k\tau \gtrsim 2\xi$ and therefore that provides a natural UV cutoff. On the other hand, although the integration converges in the IR it can be shown that the integration peaks in the region $(8\xi)^{-1}\lesssim -k\tau \lesssim 2\xi$, where we can use the approximate solutions. }
\baq
\frac{1}{2}\langle \vec{E}^2+\vec{B}^2 \rangle \simeq \frac{1}{2} \langle \vec{E}^2 \rangle&=& \frac{1}{4 \pi^2 a^4} \int d k \,  k^2  \vert A_+' \vert^2  \nn \\
&\simeq& 1.4 \times 10^{-4} \, \frac{H^4}{\xi^3} \, e^{2 \pi \xi}, \quad \txt{for} \quad \xi \gtrsim1
\label{energy}
\eaq
should always be smaller than the total energy density $\rho_t=3H^2$. Another constraint appears when the pseudo-scalar is the inflaton\footnote{In addition it would also be interesting to explore the mixing of scalar and vector perturbations to check if they given different bounds. We thank Nemanja Kaloper for pointing out this additional possible issue.}. In that case the gauge fields could potentially spoil the flatness of the potential through the term \cite{Anber:2009ua}
\baq
\langle \vec{E} \cdot \vec{B} \rangle &=& - \frac{1}{4 \pi^2 a^4} \int d k \, k^3 \, \frac{d}{d \tau} \vert A_+ \vert^2, \nonumber\\
&=& -2.4\times 10^{-4} \, \frac{H^4}{\xi^4} \, e^{2 \pi \xi}, \quad \txt{for} \quad \xi \gtrsim1
\label{flatness}
\eaq
which should be, in that case, much smaller than $f/\al \left| V'(\phi) \right|$.

\section{Estimate of loop effects} \label{est}

The enhancement of the gauge field leads also to an enhancement of the curvature and tensor perturbations through one loop effects. Here we will, for pedagogical reasons, give only a very crude estimate of these effects, since more detailed calculations has been carried out elsewhere. From the action term $\int d\tau d^3x \sqrt{-g}\mathcal{L}_{\textrm{int}}$, with $a=-1/(\tau H)$, we obtain, using Eq. (\ref{Lint0}), the vertex factor 
\beq
\mathcal{V}_s \propto \frac{1}{H^4}\xi~,
\eeq
where the two-point correlation function of $\zeta$ receives a one-loop correction with two insertions of this vertex and two propagators corresponding to $F_{\mu \nu}\til{F}^{\mu \nu}=-4\vec E\cdot \vec B$ in the loop. Since the gauge field is resonantly produced and can be treated on shell, let us assume that we can replace the propagators with the classical expectation value $\left< \vec{E}\cdot \vec{B}\right> $. 
In this way we obtain $\mathcal{P}^2= (H^2/(2\pi |\dot{\phi}|)^4$ from the two external legs, $\left< \vec{E}\cdot \vec{B}\right>^2$ from the two propagators and $(\xi/H^4)^2$ from the two vertex factors, which using Eq. (\ref{flatness}) gives, for the power spectrum of scalar curvature perturbations,
\beq\label{loop1}
\mathcal{P}_\zeta^{\textrm{one-loop}} = \gamma_s \frac{\mathcal{P}^2}{\xi^6}e^{4\pi\xi}, 
\eeq
where the precise number $\gamma_s$ needs to be computed from a more detailed evaluation, leading to \cite{Barnaby:2010vf} $\gamma_s \simeq 8\times 10^{-5}$. Similarly, by insertion of one more external leg, one more propagator, one more vertex factor, and a momentum conserving delta-function, one finds for the three-point function
\beq \label{3PF}
\left< \zeta_{k_1} \zeta_{k_2} \zeta_{k_3}\right>^{\textrm{one-loop}} = (2\pi)^3\delta^{(3)}(\sum_i\vect{k}_i)f(k_1,k_2,k_3) \frac{\mathcal{P}^3}{\xi^9}e^{6\pi\xi} , 
\eeq
where $f(k_1,k_2,k_3)$ is a momentum configuration dependent constant.  A precise calculations gives in the equilateral configuration
\beq 
f_{NL}^{\textrm{one-loop}} =\gamma_{NG} \frac{\mathcal{P}^3}{\mathcal{P}_\zeta^2\xi^9}e^{6\pi\xi}, 
\eeq
 with $\gamma_{NG} = 3\times 10^{-7}$ \cite{Barnaby:2010vf}. 

\begin{figure}
\begin{center}
\fcolorbox{white}{white}{
  \begin{picture}(451,59) (28,-12)
    \SetWidth{1.0}
    \SetColor{Black}
    \Line[arrow,arrowpos=0.5,arrowlength=5,arrowwidth=2,arrowinset=0.2](29.312,20.152)(69.615,19.541)
    \PhotonArc(97.094,20.152)(21.59,172,532){4.58}{11}
    \Line[arrow,arrowpos=0.5,arrowlength=5,arrowwidth=2,arrowinset=0.2](114.193,20.152)(157.55,19.541)
    \Line[arrow,arrowpos=0.5,arrowlength=5,arrowwidth=2,arrowinset=0.2](195.41,20.152)(238.156,20.152)
    \PhotonArc(263.804,20.152)(19.175,171,531){4.58}{10}
    \Line[arrow,arrowpos=0.5,arrowlength=5,arrowwidth=2,arrowinset=0.2](279.07,24.426)(312.046,46.41)
    \Line[arrow,arrowpos=0.5,arrowlength=5,arrowwidth=2,arrowinset=0.2](282.124,6.107)(312.046,-11.602)
    \Text(58.623,1.832)[lb]{\Black{$\nu_s$}}
    \Text(127.627,1.832)[lb]{\Black{$\nu_s$}}
    \Text(227.775,1.832)[lb]{\Black{$\nu_s$}}
    \Text(293.115,21.984)[lb]{\Black{$\nu_s$}}
    \Text(276.628,-9.771)[lb]{\Black{$\nu_s$}}
    \Gluon(350.517,19.541)(390.21,19.541){4.58}{5}
    \PhotonArc(414.636,19.541)(19.175,171,531){4.58}{10}
    \Gluon(439.062,20.152)(479.366,20.152){4.58}{5}
    \Text(382.882,1.832)[lb]{\Black{$\nu_t$}}
    \Text(443.337,1.832)[lb]{\Black{$\nu_t$}}
  \end{picture}
}
\end{center}
    \caption{Feynman diagrams associated with the one-loop contribution to the 2 and 3 point function of the scalar curvature and to the 2 point function of the tensor curvature}
\end{figure}
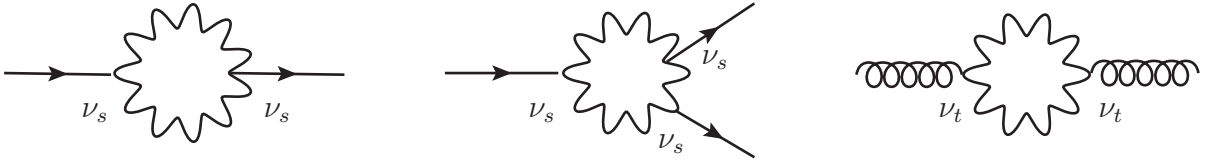

In a similar way, we can use that gravitational waves couple to the energy momentum tensor with strength $1/M_p^2$, so the vertex for the tensor modes is, in natural units,
\beq
\mathcal{V}_t \propto \frac{1}{H^4}~,
\eeq
while the propagator in the loop is dominated by the term $\left< \lmk\delta_{ij}-\partial_i\partial_j/\triangle \rmk E_i E_j \right>$ which can be roughly estimated by $\left< \vec{E}^2 \right>$. Proceeding as above we find, using Eq. (\ref{energy}),

\beq
\mathcal{P}_t^{\textrm{one-loop}} = 16 \gamma_t \frac{\left(\epsilon\mathcal{P}\right)^2}{\xi^6}e^{4\pi\xi}, 
\eeq
where a more precise calculation would give $\gamma_t \simeq 3 \times 10^{-5}$ \cite{Sorbo:2011rz}.

From this general estimates, we conclude that there could be a large effect on scalar and tensor perturbations if gauge fields are produced during inflation, which would manifest itself primarily through a very non-Gaussian contribution to the scalar perturbations. This leads to strong constraints on gauge field production during inflation in this scenario.


\section{Universal coupling of gravity} \label{inter}

In order to see the problem with decoupling the effect of the gauge field production from the curvature perturbation let us consider again the comoving gauge where there are no perturbations in the inflaton. Then, from linearized gravity, we know that, by definition, the metric fluctuations are the term in the Hamiltonian that couples linearly to the energy momentum tensor
\beq
H_{I} \propto  T^{\mu\nu} \delta g_{\mu\nu}~.
\eeq
On this form it is clear that the coupling of the metric fluctuation is democratic, in the sense that it couples universally to all species in the energy momentum tensor. Therefore, one would also not expect the detailed form of the coupling between curvature perturbations and the gauge field to depend on whether we identify the rolling axion in the coupling $\sigma F\tilde F$ with the inflaton or some other field.

In order to see how this workout in more details, it is useful to use the ADM formalism where the metric is written as
\beq
ds^2 = -N^2 dt^2 + h_{ij}(dx^i+N^i dt)(dx^j+N^jdt).
\eeq
The dynamic degrees of freedom are contained in $h_{ij}$ while the lapse ($N$) and the shift ($N^i$) are determined by the constraint equations, which can be derived by inserting the ADM decomposition of the metric into the background Lagrangian, leading to
\bea
\label{Lsigma}
\mathcal{L} = \frac{a^3}{2}\left[NR^{(3)} -2N V+ N^{-1}(E^j_i E^i_j-(E^i_i)^2)\right.\left. +\, N^{-1}(\dot\phi-N^i\partial_i\phi)^2
-N a^{-2}h^{ij}\partial_i\phi\partial_j \phi \right]~,
\eea
where $E_{ij} =\hf( \dot h_{ij} -\nabla_iN_j-\nabla_jN_i)$ is the rescaled extrinsic curvature, $V$ is the inflaton potential, and $R^{(3)}$ is the curvature scalar of the three-metric $h_{ij}$.  
We consider the gauge field as a perturbative quantitive and therefore the electromagnetic part of the Lagrangian is only relevant at quadratic order. Furthermore, we are interested on interactions between gauge fields and the comoving curvature perturbation. Therefore, in order to compute the relevant interaction at cubic order it is enough to compute the lapse and the shift to linear order. Below we compute this interaction in two different gauges.

\subsection{Comoving gauge}
In the comoving gauge, where $\delta \phi=0$ and $h_{ij} = a^2 e^{2\zeta} \left[e^{\gamma}\right]_{ij}$ \footnote{ In this gauge $\ze=-\psi$ where $\psi$ is the gravitational potential.}, the relevant interaction Hamiltonian for general matter can be written, in cosmic time, as \cite{Chaicherdsakul:2006ui, Jain:2012vm}
\baq
H_I (t)&=&\int d^3 x a^3 \, \lmk -\hf T^{\mu \nu} \delta g_{\mu \nu} + \frac{\al \delta \sigma}{4f} F_{\mu \nu} \til{F}^{\mu \nu} \rmk,
\label{Hint}
\eaq
where $T^{\mu \nu}$ is the full energy momentum tensor up to quadratic order. The second term on the interaction Hamiltonian should be included whenever the field coupled to $F_{\mu \nu} \til{F}^{\mu \nu}$ is not the inflaton but an isocurvature field ($\si$). In this gauge, $\delta \si =\dot{\si} {\cal{S}}_{\si \phi}/H$ where ${\cal{S}}_{\si \phi}$ is the gauge-invariant generalized entropy perturbation defined as
\baq
{\cal{S}}_{x y}= H \lmk \frac{ \delta x}{\dot{x}} - \frac{\delta y}{\dot{y}} \rmk. 
\eaq
The first term in Eq. (\ref{Hint}), which is the contribution to the curvature perturbation, can be simplified by noticing that \cite{Chaicherdsakul:2006ui}
\baq
-\hf a^3 T^{\mu \nu} \delta g_{\mu \nu}=a^3 \lkk \frac{\dot{\ze}}{H}T^{00}-\partial_i\lmk -\frac{\ze}{H}+\e a^2 \nabla^{-2}\dot{\ze} \rmk T^{0i}-a^2 \ze T^{ii} \rkk
\eaq
and using the identity \cite{Chaicherdsakul:2006ui}
\baq
\nabla_\mu T^{\mu 0}&=&a^{-3} \partial_t \lmk a^3 T^{00} \rmk + a \dot{a} T^{ii}+ \partial_i T^{i0}.
\eaq
Therefore, the interaction with curvature perturbations term can be rewritten as
\baq
-a^3\int d^3 x \hf \, T^{\mu \nu} \delta g_{\mu \nu}&=&\int d^3 x \lkk-  a^3 \frac{\ze}{H} \nabla_\mu T^{\mu 0} + \frac{1}{H} \partial_t \lmk  a^3 \ze T^{0 0} \rmk \rkk
\eaq
up to leading order in slow-roll. The total time derivative can be canceled by a field redefinition. Thus, we just need to compute the quantity $\nabla_\mu T^{\mu 0}$ where the relevant energy momentum tensor up to quadratic order is simply\footnote{The axial coupling does not contribute to the energy momentum tensor as $\delta \left( \sqrt{-g} F_{\mu \nu} \tilde{F}^{\mu \nu} \right) / \delta g^{\mu \nu} = \delta \left( 1/2 F_{\mu \nu} \e^{\mu \nu \al \be} F_{\al \be} \right)/ \delta g^{\mu \nu} =0$ } 
\baq
T^{\mu \nu}= F^{\mu \be}F^{\nu}_{\ \be}  -\frac{1}{4} g^{\mu \nu} F_{\al \be}  F^{\al \be}.
\eaq
The conservation of momentum only holds for the total energy momentum tensor and not for this part in isolation, hence, $\nabla_\mu T^{\mu 0} \neq 0$. Using Eq. (\ref{Aeom0}) and the Bianchi identity we find that
\baq
\nabla_\mu T^{\mu \nu}&=& F^\nu_{\, \be} \nam F^{\mu \be} + F_{\mu \be} \nabla^\mu F^{\nu \be}-\hf F_{\al \be}\nab^{\nu} F^{\al \be}  \nn \\
&=& - \frac{\al}{f} F^\nu_{\, \be} \, \til{F}^{\mu \be} \, \nam \si.
\eaq
Therefore, if we assume $\si$ to be an homogeneous field,
\baq
\nabla_\mu T^{\mu 0}=-\frac{ \al \dot{\si}}{f} \vec{E} \cdot \vec{B}.
\eaq
Finally, using the fact that $F_{\mu \nu} \til{F}^{\mu \nu}=-4 \vec{E} \cdot \vec{B}$ we arrive at the interaction Hamiltonian which, to leading order, gives
\baq
H_I( \tau \pr)&=&- \frac{ \al }{f} \int d^3 x \frac{a^3\dot{\si}}{H} \lmk -\ze +  {\cal{S}}_{\si \phi} \rmk \vec{E} \cdot \vec{B} =  -2 \xi  a^3 \int d^3x   \lmk {\cal R} +  {\cal{S}}_{\si \phi} \rmk \vec{E} \cdot \vec{B}
\label{Hintf}
\eaq
where ${\cal R}$ is the gauge-invariant comoving curvature perturbation\footnote{In fact $\ze=-\psi\simeq -{\cal{R}}+\frac{\dot{\si}^2}{\dot{\phi}^2} {\cal S}_{\si \phi}$. However, as the curvature perturbation is mainly adiabatic, $\dot{\si} \ll \dot{\phi}$, and therefore the second term should be subdominant.}.

\subsection{Spatially flat gauge}
Let us do the same approach as before but now in the spatially flat gauge where $\psi=0$ and $h_{ij} = a^2 \left[e^{\gamma}\right]_{ij}$. In this case, the leading contribution at cubic order to the interaction Lagrangian between curvature perturbations and the gauge fields is \cite{Barnaby:2011vw}\footnote{As mentioned in \cite{Barnaby:2011vw} there are other cubic contributions coming from the lapse function but they are parametrically $\sqrt{ \e}$ smaller than this term.}
\baq
\label{S3}
  S= \frac{\alpha}{f} \int dt d^3 x \, \lmk a^3 \delta \si\,  \vec{E} \cdot \vec{B}  \rmk.
\eaq
If $\si$ is the inflaton it is straightforward to get the leading interaction with gauge fields, ${\cal{L}}_\txt{int} = \al \dot{\si}/(f H) \ze\vec{E} \cdot \vec{B}$. If an isocurvature field couples to the EM sector then there is a term in the interaction Hamiltonian equal to the second term in Eq. (\ref{Hint}).
In the gauge where $\psi=0$ we identified this term as an isocurvature contribution to the curvature perturbation. The same is not true is this gauge. Namely, if we use the two gauge invariant quantities ${\cal{R}}$ and ${\cal{S}}_{\si \phi}$, which in the spatially flat gauge are given by \cite{Gordon:2000hv}
\baq
&&{\cal R}=H \frac{\delta \phi}{\dot{\phi}} \quad \txt{and} \quad {\cal{S}}_{\si \phi}= H \lmk \frac{ \delta \si}{\dot{\si}} - \frac{\delta \phi}{\dot{\phi}} \rmk,
\eaq
we can rewrite $\delta \si$ in terms of these two quantities as
\baq
\quad \delta \si=\frac{\dot{\si}}{H} \lmk {\cal{S}}_{\si \phi}+ {\cal R} \rmk.
\eaq
Therefore, in this gauge the interaction Hamiltonian becomes
\baq
\label{S4}
  S&=& \frac{\alpha}{f} \int dt d^3 x \, \lmk a^3 \delta \si\,  \vec{E} \cdot \vec{B}  \rmk=  \frac{\alpha}{f} \int dt d^3 x \, \frac{a^3\dot{\si}}{H} \,\lmk {\cal R} +{\cal{S}}_{\si \phi} \rmk \vec{E} \cdot \vec{B}  \nn \\ 
  &=&  2 \xi   \int dt d^3x \,  a^3\lmk {\cal R} + {\cal{S}}_{\si \phi} \rmk \vec{E} \cdot \vec{B} 
\eaq
which is in agreement with Eq. (\ref{Hintf}).
We verify then that the contribution to the curvature perturbation should be exactly the same as a function of the interaction coefficient $\xi$, independently of which pseudo-scalar field is coupled to $F \til{F}$.

Of course one would expect that in the limit where $\sigma$ is a pure isocurvature field, the perturbations of the $\sigma$-field will decouple from the curvature perturbation. This is the limit where the energy density in $\sigma$ vanishes, which implies $\dot\sigma \to 0$. Indeed it is true that the coupling of the curvature perturbation with the gauge field in (\ref{S3}) vanishes in this limit, where also $\xi$ vanishes. However, this case is quite trivial as there is no gauge field production either. One could also consider the case where $\dot\si\to 0$ on super-horizon scales. This scenario, mentioned in \cite{Barnaby:2012xt}, seems to be a caveat to the previous reasoning, although, as we show in next section, also in this case there are non-trivial constraints which need to be satisfied in order for the final total curvature perturbation to be suppressed.

\section{Superhorizon evolution of the curvature perturbation} \label{SHevSec}

In this section we will analyze the superhorizon evolution of curvature perturbation. The analysis can be divided in two different cases depending on whether the $\si$ field decays before or after the end of inflation. 

The curvature perturbation remains roughly constant during inflation if the field does not decay. The same happens during radiation domination if the isocurvature field decays quickly into radiation. If not, the field will become energetically more relevant which will increase the of total value of curvature perturbation. Instead, if the field decays into cold dark matter then isocurvature perturbations, which are strongly constrained by Planck observations \cite{Ade:2013ydc}, are generated in addition. Therefore, if the field only decays after reheating we can work under the conservative assumption that the curvature perturbation remains at least constant at late times.

The other scenario is more delicate. In the case where the field decays during inflation, curvature perturbation is similarly generated at horizon crossing through the coupling with $F \til{F}$, but its contribution to the total curvature perturbation is erased as soon as the isocurvature field decays and only the inflaton perturbation remains \cite{Mukohyama:2014gba}. The inflaton perturbation is only coupled gravitationally to the gauge fields and therefore the final sourcing of the total curvature perturbation becomes slow-roll suppressed compared to the other scenarios. Nevertheless, even in  this case there are some non-trivial constraints coming from the slow-roll variation of $\cR_\phi$ on superhorizon scales due to the relation \cite{Linde:2005he}
\beq
 \cR_\phi'=-\lmk \frac{\dot{\si}}{\dot{\phi}} \rmk^2  \cR_\sigma'.
\eeq
In conformal time $\tau$, it gives
\baq \label{SHev}
\cR_\phi(\tau)= \cR_\phi^*- \int^{\tau_{f}}_{\tau^*} \lmk \frac{\dot{\si}}{\dot{\phi}} \rmk^2 \cR_\sigma' \, d\tau ,
\eaq
where the star denotes quantities evaluated at horizon crossing or, more precisely, as soon as the source shuts down which is around the time of horizon crossing. In the case of $\cR_\phi$ we assume it to be dominated by the vacumm contribution at those times. The integration in the right hand side of Eq. (\ref{SHev}) can be divided in two stages, the first one where $\dot \sigma \simeq \dot \sigma_0$ is constant and a second stage where the $\sigma$ becomes heavy, starts oscillating and decays as $\dot \si = \dot \si_0 \cos (m_\sigma (t-t_{osc})) (a_{osc}/a)^{3/2}$. The only unknown in the integrand is the value of  $\cR_\si'$. One could expect that it would be proportional to the slow-roll parameters of $\si$, but that is not the case due to the mixing of $\sigma$ with the inflaton $\phi$ through the non-diagonal mass mixing term of the form \cite{Sasaki:1986hm, Mukhanov:1988jd}
\beq\label{Ldpds}
\mathcal{L}_{\delta\phi\delta\si} \supset 3H \dot\phi\dot\si \delta\phi\delta\si~.
\eeq
Generically, if we have a set of scalar fields, $\phi_I$, during inflation, their quantum fluctuations, $\pI$, will satisfy the following equations of motion
\baq \label{Mixeom1}
\ddpI + 3H \dpI + \frac{k^2}{a^2} \pI + \sum_J \lkk V_{IJ} - \frac{1}{a^3} \frac{d}{dt} \lmk \frac{a^3}{H} \dot{\phi_I} \dot{\phi_J} \rmk \rkk \delta \phi_J=S_I ,
\label{eoma}
\eaq
where the source term is given by 
\baq
S_I=
\begin{pmatrix}
0 \\
\frac{\al}{f} \vec{E} \cdot \vec{B}
\end{pmatrix},
\eaq
and $V_I=\p V/\p \phi_I$.
In our case the fields are the inflaton, $\phi$, and an isocurvature field, $\s$, which are only gravitationally coupled, hence we have $V_{\phi \s}=0$. The equations of motion can be simplified by defining $u_I=a \delta \phi_I$ and by working in conformal time $\tau=-(1+\e)/(a H)$. This way Eq. (\ref{Mixeom1}) becomes \cite{Byrnes:2006fr}
\baq
u_I''+ \lmk k^2 - \frac{2}{\tau^2} \rmk u_I- \frac{3}{\tau^2} \sum M_{IJ} u_J=a^3 S_I,
\eaq
where $M_{IJ}$ contains the mixing terms and it is given, at first order in slow-roll, by
\baq
M_{IJ} \simeq 
\begin{pmatrix}
3\e_\phi-\eta_\phi &  2 \theta \e_\phi \\
2 \theta \e_\phi & \e_\phi-\eta_\si 
\end{pmatrix},
\eaq
where we have defined $\theta \equiv \dot \sigma / \dot \phi \ll 1$.
If both fields are in slow-roll, which is the case in the first stage, the mixing matrix is approximately constant and we can obtain the eigenstates $\{v_1, v_2 \}$ before $\si$ starts to oscillate by diagonalizing $M_{IJ}$, namely, by defining $U_{IJ}$ such that
\baq
U_{IJ} \, v_J = u_I \quad \txt{and} \quad U^T M U = \txt{diag} (\lambda_1, \lambda_2),
\eaq
where 
\baq
\lambda_{1,2}\simeq \hf \lmk 4 \e_\phi -  \eta_\phi + \eta_\si \pm | 2 \e_\phi - \eta_\phi + \eta_\si | \rmk
\eaq
are the eigenvalues of $M_{IJ}$. Then, $v_I$ satisfies the decoupled system of equations
\baq
v_I''+ \lkk k^2 - \frac{1}{\tau^2} \lmk \mu_I^2-\frac{1}{4} \rmk \rkk v_I=a^3 U_{JI}^T S_J \equiv \til{S}_I,
\eaq
where we have defined $\mu_I \equiv 3/2 + \lambda_I$. The matrix $U_{IJ}$ is a rotation matrix of angle $\Theta$ such that \cite{Byrnes:2006fr}
\baq
\tan 2\Theta= \frac{ 4 \theta \e_\phi }{ 2 \e_\phi - \eta_\phi + \eta_\si}.
\eaq
Under our assumptions $\Theta \ll 1$ and therefore $U_{IJ}$ can be written as
\baq
U_{IJ}\simeq
\begin{pmatrix}
1 & - \Theta   \\
  \Theta  &  1
\end{pmatrix} \label{rotm}.
\eaq
Note that the source term is also rotated by $U_{IJ}^T$  to
\baq
\til{S}_I = a^3 \frac{\al}{f} \vec{E} \cdot \vec{B} 
\begin{pmatrix}
 \Theta  \\
1
\end{pmatrix},
\eaq
hence, it will have a much stronger effect on the eigenstate $v_2$ than on $v_1$.

While the fields are slowly rolling, the homogeneous solutions for the mass eigenstates, $v_I^0$, are the standard ones for quantum fluctuations of massive fields, which are approximately constant solutions at superhorizon up to slow-roll corrections embedded in $\lambda_I$. The particular solutions, $v_I^p$ of the equation of motion are directly related to the case where the inflaton is coupled to $F\til{F}$, which is defined as $\delta \phi^{inv. dec.}$ in \cite{Barnaby:2011vw}. In particular, as the source term only acts around the time of horizon crossing we can write
\baq
\begin{pmatrix}
\lb v_1^p \rb   \\
\lb v_2^p \rb 
\end{pmatrix}
= a \lb \delta \phi^{inv. dec.} \rb 
\begin{pmatrix}
\Theta   \\
1
\end{pmatrix}.
\label{invdec}
\eaq 
If we now come back to the $\{ \delta \phi, \delta \si \}$ basis we have
\baq
a \, \delta \phi&=& v_1 -  \Theta   \,v_2   \nn \\ 
a \, \delta \s&=&   \Theta  \, v_1 +  \,  v_2  \,. \label{decom}
\eaq
In what concerns $\delta \si$, we assume that $v_2^p \gg v_2^0$ so we can safely say that $\delta \si \simeq  v_2/a$.  On superhorizon scales ($-k\tau \to 0$), the solution for $v_2$ will be
\baq
v_2 \simeq \frac{C}{\sqrt{k}} (-k \tau)^{-1-\lambda_2}, 
\eaq
where $C$ is an integration constant which has to be fixed once the source is turned off.
Therefore, the time variation of $\cR_\si$ at superhorizon scales during the first stage is given by
\baq
\cR_\si ' (\tau^*<\tau<\tau_{osc}) &=& \frac{d}{d\tau} \lmk H \frac{ \delta \si}{\dot \si_0} \rmk =\cR_\si^* \lmk \frac{ H'}{H} + \frac{  \delta\si '}{\delta \si} \rmk \nn \\
&\simeq& -a H \cR_\si^* \lmk  2  \, \e_\phi - \lambda_2 \rmk.
\eaq
In the second stage $\delta \si$ and $\si$ satisfy the same equation of motion therefore $d(\delta \si / \dot \si)/d\tau=0$. In this stage the variation of curvature perturbations will be given by
\baq
\cR_\si ' ( \tau_{osc}< \tau)&=& - \e_\phi \,a H \cR_\si^*.
\eaq
So, now we are able to compute the integral in Eq. (\ref{SHev}) until the end of inflation which yields
\baq \label{SHev2}
 \int^{\tau_{f}}_{\tau^*} \lmk \frac{\dot{\si}}{\dot{\phi}} \rmk^2 \cR_\sigma' \, d\tau &= &\int^{\tau_{osc}}_{\tau^*} \lmk \frac{\dot{\si}}{\dot{\phi}} \rmk^2 \cR_\sigma' \, d\tau + \int^{\tau_{f}}_{\tau_{osc}} \lmk \frac{\dot{\si}}{\dot{\phi}} \rmk^2 \cR_\sigma' \, d\tau \nn \\
 &\simeq&  \lmk \frac{\dot{\si}_0}{\dot{\phi}} \rmk^2  \cR_\si^* \lkk \log \lmk \frac{\tau_{osc}}{\tau^*} \rmk \lmk  2 \e_\phi - \lambda_2 \rmk- \e_\phi  \int^{\tau_{f}}_{\tau_{osc}}  \lmk \frac{\dot \sigma}{\dot \si_0} \rmk^2 \, \frac{d \tau}{\tau} \rkk,
\eaq
As $ \log \lmk \tau^*/\tau_{osc} \rmk = \Delta N$ is the duration of the first stage in e-folds we get
\baq \label{SHev3}
 \int^{\tau_{f}}_{\tau^*} \lmk \frac{\dot{\si}}{\dot{\phi}} \rmk^2 \cR_\sigma' \, d\tau &\simeq& - \lmk \frac{\dot{\si}_0}{\dot{\phi}} \rmk^2  \cR_\si^* \lkk \Delta N  \lmk  2 \e_\phi - \lambda_2 \rmk+\frac{\e_\phi}{6}  \rkk
 \eaq 
Finally, using Eq. (\ref{SHev}) we obtain the value of $\cR_\phi$ at the end of inflation which is given by
\baq
\cR_\phi(\tau)=\cR_0+ \lmk \frac{\dot{\si}_0}{\dot{\phi}} \rmk^2   \cR_\sigma^* \lkk \Delta N  \lmk  2 \e_\phi - \lambda_2 \rmk +\frac{\e_\phi}{6}    \rkk.
\eaq
For example, in the case of chaotic inflation with a quadratic potential and assuming $\eta_\si \ll \e_\phi $ the eigenvalue gives $\lambda_2= \e_\phi$ which means that 
\baq \label{SH5}
\cR_\phi(\tau) \gtrsim \Delta N \, \e_\phi \lmk \frac{\dot{\si}_0}{\dot{\phi}} \rmk^2   \cR_\sigma^*. 
\eaq
This inequality remains true for any other inflationary model unless there is a precise cancelation of the $2 \e_\phi - \lambda_2$ term. In particular, it becomes a very conservative lower limit when $\eta_\phi \gg \e_\phi$.
Although the sourcing of $\cR_\phi$ is parametrically suppressed by $\e_\phi$ the slow-roll logarithmic corrections give a term proportional to duration of the first stage in e-folds. If the first stage is long enough this term gives a very important contribution to the curvature perturbation. Even if $\Delta N=1$ this contribution is still larger than the direct sourcing of $\cR_\phi$ at horizon crossing. Thus, when the $\sigma$-field has decayed and the total curvature perturbation is given simply by $\cR \approx \cR_\phi$ we can write
\beq \label{SRPS}
P_{\zeta}\simeq {\cal P} \lmk 1+ \gamma_s\,  \Delta N^2 \e^2 \frac{ {\cal P}}{\xi^6} e^{4\pi \xi} \rmk~.
\eeq
If we compare with the effect on the tensor spectrum, which does not evolve on super horizon scales
\beq
P_{{\rm GW}} \simeq  16  \e  {\cal P} \lmk 1+ \gamma_t\, \frac{\e  {\cal P}}{\xi^6} e^{4\pi \xi} \rmk,
\eeq
we see that if $\Delta N^2 \gtrsim 16 \, \gamma_t/\gamma_s$ the non-Gaussian contribution to the scalar perturbation becomes larger than the tensor power spectrum. Thus, if one has sensitivity to observe the tensor modes, one would also expect to have sensitivity to observe the non-Gaussian scalar modes. Inserting the numbers, we see that the axion would have to decay within two e-folds, $\Delta N \approx 2.4$, in order not to source a non-Gaussian contribution to the spectrum larger than the tensor spectrum. 

Similarly, the same effect is important when computing the 3-point function. The result can be easily extrapolated because $\left< \zeta_\phi \zeta_\phi \zeta_\phi \right> \gtrsim \e^3 \Delta N^3 \left< \zeta_\si \zeta_\si\zeta_\si\right>$ where the 3-point function of $\zeta_\si$ is the same as the 3-point function in the case of axial coupling with the inflaton, given in Eq. (\ref{3PF}).

\section{The implications} \label{impli}

The universal nature of the coupling between curvature perturbation and the gauge fields, when written in terms of $\xi$, leads to the very generic result in Eq. (\ref{loop1}) independent of which axion is contributing to $\xi$. In particular, if the observable CMB anisotropies come from the curvature perturbation generated during inflation, then observations tells us $\mathcal{P}\simeq 10^{-9}$. Requiring then $\mathcal{P}_\zeta^{\textrm{one-loop}} \ll \mathcal{P}$, implies the generic constraint\footnote{In the minimal scenario where we can assume by simple extrapolation that the dynamics of inflation is known all the way from when the observable modes exit the horizon until the end of inflation, then, due to the fact that the coupling $\xi$ increases in time during inflation, more stringent constraints on $\xi$ were derived in the case where $\si$ is the inflaton, based on departures from scale invariance \cite{Meerburg:2012id} and the formation of primordial black holes \cite{Linde:2012bt}. We believe that these bounds can similarly be generalized to the case where $\si$ is not the inflaton due the universal nature of the coupling between the curvature perturbation and the gauge fields discussed above in the previous sections.} 
\beq\label{constr1}
\xi\lesssim 3~,
\eeq
even if the axion is not identified with the inflaton. Only if the field decays during inflation or the observed perturbations are generated later, like in the curvaton mechanism, such that $\mathcal{P}\ll 10^{-9}$, the constraint (\ref{constr1}) can be relaxed. We discuss some of these different situations below.

\subsection{Chiral Gravitational Waves}

In the previous section it was shown that the interaction Lagrangian with adiabatic perturbations remains exactly the same whether the pseudo-scalar coupled to $F\til{F}$ is the inflaton or not. Therefore, the same constraints should apply in both cases, namely, the results derived in \cite{Barnaby:2010vf, Sorbo:2011rz} for the power spectrum of scalar and tensor perturbations
\baq \label{PS0}
P_{\zeta}&\simeq& {\cal P} \lmk 1+ \gamma_s\, \frac{ {\cal P}}{\xi^6} e^{4\pi \xi} \rmk, \\
P_{{\rm GW}} &\simeq&  16  \e  {\cal P} \lmk 1+ \gamma_t\, \frac{\e  {\cal P}}{\xi^6} e^{4\pi \xi} \rmk,
\eaq
where $\gamma_{s,t}$ are some numerical coefficients, ${\cal P}^{1/2}= H^2/(2\pi |\dot{\phi}|)$ and we depreciated the mild $k$ dependence. This fact has strong consequences for the production of gravitational waves by this mechanism. Namely, the tensor contribution to the power spectrum is parametrically smaller by a factor of $\e$ than the scalar contribution, when one compares both to the vacuum production. Therefore, the only way gravitational waves could be generated without spoiling the power-spectrum would be if $\gamma_t \gg \gamma_s$, which is not the case. Thus, the generation of gravitational waves by this mechanism is smaller than the vacuum contribution.
Even if a curvaton generates the spectrum of scalar perturbations \cite{Enqvist:2001zp, Lyth:2001nq, Moroi:2001ct}\footnote{See also \cite{Mollerach:1989hu,Linde:1996gt} for some early related developments prior to the curvaton mechanism put forward in \cite{Enqvist:2001zp, Lyth:2001nq, Moroi:2001ct}.} the tensor to scalar ratio ($r$) would be given by
\baq
r=\frac{16 \gamma_t\, \e^2  {\cal P}^2 \frac{e^{4\pi \xi}}{\xi^6} }{ P_\ze^\txt{obs}}.
\eaq
However, the contribution for the scalar spectrum should still satisfy 
\baq
 \gamma_s\, \frac{ {\cal P}^2}{\xi^6} e^{4\pi \xi} < P_\ze^\txt{obs} \simeq 2 \times 10^{-9} \quad \Ra \quad r < 16 \frac{\gamma_t}{\gamma_s} \e^2 \simeq 6 \, \e^2~.
 \eaq
An even stronger constraint appears from non-Gaussianity. In \cite{Barnaby:2010vf} non-Gaussianity in the scalar power spectrum was computed and shown to peak in the equilateral shape. The associated non-Gaussian parameter $f_\txt{NL}^\txt{eq}$ parameter was computed to be \cite{Barnaby:2010vf}
\baq
f_{\txt{NL},\si}^\txt{eq} \simeq \gamma_{NG} \frac{ {\cal P}^3}{\lmk P_\ze^\txt{obs}\rmk^2} \frac{e^{6 \pi \xi}}{\xi^9 }, \quad \gamma_{NG}=3 \times 10^{-7}.
\eaq
The current constraint is $f_\txt{NL}^\txt{eq} = -42 \pm 75$ \cite{Ade:2013ydc}. Thus, this requires that
\baq
f_{\txt{NL},\si}^\txt{eq}< f_\txt{NL}^\txt{eq} \quad &\Ra& \quad \frac{ {\cal P}^2}{\xi^6} e^{4\pi \xi} <  \lmk \frac{f_\txt{NL}^\txt{eq} \lmk P_\ze^\txt{obs}\rmk^2}{\gamma_{NG}} \rmk^{2/3}, 
\eaq
which then implies that the tensor to scalar ratio has to satisfy the upper bound
\baq\label{r0}
r<16\, \gamma_t\,  \e^2  \lmk \frac{ f_\txt{NL}^\txt{eq}  }{\gamma_{NG}} \rmk^{2/3} \lmk P_\ze^\txt{obs} \rmk^{1/3} \simeq   10^{-2} \e^2   \lmk f_\txt{NL}^\txt{eq}  \rmk^{2/3}. 
\eaq
Therefore, in order for the tensor modes generated through this mechanism to overcome the vacuum production tensor, the tensor to scalar ratio cannot be larger than $r_\txt{max}\simeq {\cal O} \lmk 10^{-5} \rmk$. This value is far below the recent claimed observation of primordial tensor modes from BICEP2 \cite{Ade:2014xna} but also below any near-future observation.

Finally we look at the case where the field $\si$ decays during inflation. As we verified in the last section, although the curvature perturbation is generated at horizon crossing, it is later erased when the $\sigma$ field decays. Thus, in the end of inflation $\cR \simeq \cR_\phi$ where the source effect in $\cR_\phi$ is slow-roll suppressed when compared to the one in $\cR_\si$. Nevertheless, even in this case and due to the relation between $\cR_\phi$ and $\cR_\si$ on super horizon scales, we derived in Eq. (\ref{SH5}) that there is a superhorizon enhancement proportional to the number of e-folds since horizon-crossing of the mode until the decay of the field.  In such a case the scalar power spectrum is modified to
\baq
P_{\zeta}&\simeq& {\cal P} \lmk 1+  \gamma_{s}\, \Delta N^2 \e^2 \frac{ {\cal P}}{\xi^6} e^{4\pi \xi} \rmk.
\eaq
Therefore, although the contribution to the power spectrum is $\e^2$ suppressed, the $\gamma_s \, \Delta N^2$ factor gives a significant contribution. Similarly, the associated non-Gaussian parameter now becomes
\baq
f_{\txt{NL},\si}^\txt{eq} \simeq  \e^3 \Delta N^3\, \gamma_{NG}  \frac{ {\cal P}^3}{\lmk P_\ze^\txt{obs}\rmk^2} \frac{e^{6 \pi \xi}}{\xi^9 }.
\eaq
Even if $\Delta N\simeq 1$ the non-Gaussian parameter derived here is 2 orders of magnitud larger than the one associated only with the sourcing of $\cR_\phi$ at horizon crossing \cite{Barnaby:2012xt}. Imposing $f_{\txt{NL},\si}^\txt{eq}< f_\txt{NL}^\txt{eq}$ implies the following upper bound
\baq\label{r1}
r\lesssim \frac{10^{-2}}{\Delta N^2}   \lmk f_\txt{NL}^\txt{eq}  \rmk^{2/3}.
\eaq
The new contribution to the curvature perturbation derived here in the case where the $\si$ field decays during inflation has important implications for the models proposed in \cite{Cook:2013xea, Mukohyama:2014gba, Caprini:2014mja} where large CMB tensor modes, larger than the vacuum ones, can be generated through this mechanism. 

While our constraint in Eq. (\ref{r1}) is an actual constraint, we can also compare it with the forecast for the parameter $X\equiv \e \,e^{2 \pi \xi}/\xi^3 < 5\cdot 10^5$, derived in \cite{Shiraishi:2013kxa} from an analysis on the tensor bispectrum signatures on the CMB due to the referred mechanism. Our constraint becomes stronger than the forecasted constraint whenever $\Delta N \gtrsim 1.2$. Nevertheless, a tensor to scalar ratio $r \simeq {\cal O} (10^{-1})$ is still allowed if the field decays faster than approximately $1.7$ e-folds. 

\subsection{Axion decay constants and Natural Inflation}

Another important cosmological consequence of this analysis shows up in the context of inflationary theories where axions play a role, like natural inflation \cite{Freese:1990rb} or its UV completed versions \cite{Kim:2004rp, Dimopoulos:2005ac}\footnote{See \cite{Kenton:2014gma, Harigaya:2014ola, Abe:2014pwa, Higaki:2014mwa, Ben-Dayan:2014lca, Long:2014dta, Kappl:2014lra} for some related recent ideas.}. Even though some of these proposals generate a super-Planckian effective decay rate, one should be careful about the axial coupling to gauge fields. That can be problematic because, as already mentioned in \cite{Barnaby:2010vf, Barnaby:2011vw} , generically, each axion ($\si_i$) couples with a given gauge field(s) as $\al_i \si_i/ f_i F \til{F}$ where $\al_i=g_i^2/(2 \gamma_i)$ and $g, \gamma_i$ are, respectively, the coupling constant to the gauge field and some numerical coefficient depending on the precise coupling to fermions. Therefore, it is not expected that we can rearrange all the axial couplings as $\Phi/f_\Phi F\til{F}$ where $\Phi$ is a collective state of axions with an effective super-Planckian decay constant $f_\Phi$. Thus, in general, we need to deal with each term separately. This is possibly interesting because, as derived in \cite{Barnaby:2010vf}, non-Gaussian constraints imply $\xi \lesssim 3$. This constraint should be satisfied by all axial couplings
\baq
\xi_i \lesssim 3 \quad \Ra \quad f_i \gtrsim \frac{\al_i}{3} \sqrt{\frac{\e_i}{2}} \,M_p,
\eaq
where we defined a first slow-roll parameter $\e_i$ associated with each axion. For example, if the axial coupling is with the Standard Model $U(1)$ gauge field, then, $g_i^2/(4 \pi)=1/137$, and for $\gamma_i \simeq1$ we get the constraint $f_i \gtrsim  5 \times10^{-2} \sqrt{\e_i}\, M_p$. In some specific scenarios where the number of axions is very large, like N-flation, the constraint on $f_i$ could be even stronger if all the axions couple to the same gauge field because then the condition $\sum_{i=1}^N \xi_i <3$ should be satisfied. For instance, if we assume, for simplicity, $\al_i=\al$, $\e_i=\e$ and $f_i=f$, then the constraint becomes $  f \gtrsim N \al \sqrt{\e} \,M_p$ in this particular scenario. 

In short, this is a very non-trivial lower bound which makes the theory constrained both from above and from below.

\subsection{The axion as a curvaton}

As a final example, let us consider the case where the curvaton is an axion with potential
\beq
V(\sigma) = \Lambda^4\left[1-\cos(\sigma/f_1)\right]~.
\eeq
We will, for simplicity, consider the case where all the observable curvature perturbation comes from the curvaton fluctuation. The spectral index of the curvaton is given by
\beq
n_\sigma -1 = -2\ep + \frac{2 V''(\sigma)}{3H^2}~.
\eeq
In the pure curvaton case, the scale of inflation is typically low and $\ep \ll 1$. We will, therefore, neglect $\epsilon$, and in order to have a red spectrum $n_\sigma-1<0$ consistent with observations, we must require
\beq
V''(\sigma) = m_\sigma^2 \cos(\sigma/f_1) \lesssim 0~,
\eeq
where $m_\si^2\equiv \Lambda^4/f_1^2$. This implies that we can assume $\si\simeq f_1$. Assuming that the axion couples to two different gauge groups with two different decay rates\footnote{Curvaton models, where the curvaton couples with different decay rates to two different gauge fields have been considered in \cite{Takahashi:2013tj,Sloth:2014sga}} as $(\alpha_i \si / f_i) F_i\tilde F_i$ with
\beq
f_1 > f_2 ~,
\eeq
then approximate Gaussianity of the curvaton perturbation require $\xi_2 \lesssim 1$, which implies
\beq
f_2 \gtrsim \frac{\alpha_2 \dot\sigma}{2H},
\eeq 
while from the equation of motion in the slow-roll approximation we have
\beq
3H\dot\si \approx -V'(\si) = -m_\si^2 f_1\sin(\si/f_1)~.
\eeq
Thus, assuming $\dot\si/H \simeq (m_\si^2/H^2)f_1$, we obtain an interesting constraint on the second decay constant
\beq
f_2  \gtrsim \frac{\al_2}{2} \frac{m_\si^2}{H^2} f_1~.
\eeq

\section{Conclusion} \label{conc}

In this work we discussed the role of the pseudo-scalars during inflation as possible sources of cosmological signatures due to axial coupling(s) with $U(1)$ field(s). If the inflaton, $\phi$, is itself a pseudo-scalar the sourcing of adiabatic scalar curvature perturbations puts non-trivial constraints on the parameter $\xi= \al \dot{\phi} /(2 H f)$, where $f$ is the decay rate of the inflaton. Moreover, the same axial coupling can source gravitational waves. 
This has been the focus of several recent papers \cite{Barnaby:2012xt, Cook:2013xea, Mukohyama:2014gba,Caprini:2014mja} where it was generally believed that if the pseudo-scalar in the axial coupling is instead identified with an isocurvature field, then, scalar perturbations would be suppressed by an $\e^2$ factor when compared to the inflaton case. As the gravitational wave production remains unaltered, this has been considered an exciting new possibility of having a primordial gravitational wave spectrum dominated by the axial coupling contribution without spoiling the CMB anisotropy constraints.

Here we proved in two different gauges that even if the pseudo-scalar is not identified with the inflaton, the interaction Lagrangian between curvature perturbations and gauge fields is exactly the same as in the inflaton case and, therefore, the constraints should apply in the same way in both cases.  This has implications, for example, for the maximal amount of tensor modes generated by this mechanism, as seen from our Eq. (\ref{r0}), which implies $r \lesssim 0.01 \ep^2 (f_\txt{NL}^\txt{eq})^{2/3} \simeq 0.2 \ep^2$ with the present constraint on  $f_\txt{NL}^\txt{eq}$ inserted.  A possible caveat to these constraints is the case where the field decays during inflation, as also mentioned in \cite{Mukohyama:2014gba}. However, even in this case we have demonstrated that there is a superhorizon enhancement of the curvature perturbation proportional to the number of e-folds between horizon crossing and the decay of the isocurvature field. This enhancement still severely constraints this alternative mechanism of generating gravitational waves (synthetic tensor modes) because a large tensor perturbation is likely to be associated with a large non-Gaussian signal in the scalar perturbations. Namely, from our Eq. (\ref{r1}) we have $r \lesssim 0.01 (f_\txt{NL}^\txt{eq})^{2/3} /\Delta N^2$, which implies that a large tensor-to-scalar ratio of order $r\simeq 0.1$ would require the axion to decay within $2$ e-folds of super-horizon evolution ($\Delta N\lesssim 2$).  Another possible way to circumvent the generic constraints derived in this paper is if we combine the mechanism with a curvaton scenario. In such a case the constraint from non-Gaussianity implies the tensor to scalar ratio to be smaller than ${\cal O} \lmk 10^{-5} \rmk$.

There are other interesting consequences of this analysis, for instance in the context of natural inflation and their UV complete generalizations. The axial coupling of each axion to gauge field(s) could potential lead to observable non-Gaussianity. Therefore, one should apply the non-Gaussian constraint separately to each axial coupling. This constraint turns out to be non-trivial because it puts a lower bound, $f_i \gtrsim \al_i \sqrt{\e_i} \,M_p$, on the decay rate of the axion, where $\e_i$ the is the first slow-roll parameter associated with each axion and $\al_i$ is a parameter related with the coupling of the axion to the gauge field.

\subsection*{Acknowledgements} 

We would like to thank Guido D'Amico, Nemanja Kaloper, Antonio Riotto and David Wands for their comments on the draft. In addition we would like to thank Shinji Mukohyama, Ryo Namba, Marco Peloso, Gary Shiu, and Lorenzo Sorbo for clarifying comments about their scenarios and discussions.  We would also like to thank the Lundbeck foundation for financial support.

\bibliographystyle{JHEP}
\bibliography{ChiralGW.bib}

\providecommand{\href}[2]{#2}\begingroup\raggedright\begin{thebibliography}{10}

\bibitem{Freese:1990rb}
K.~Freese, J.~A. Frieman, and A.~V. Olinto, {\it {Natural inflation with pseudo
  - Nambu-Goldstone bosons}},  {\em Phys.Rev.Lett.} {\bf 65} (1990) 3233--3236.

\bibitem{Adams:1992bn}
F.~C. Adams, J.~R. Bond, K.~Freese, J.~A. Frieman, and A.~V. Olinto, {\it
  {Natural inflation: Particle physics models, power law spectra for large
  scale structure, and constraints from COBE}},  {\em Phys.Rev.} {\bf D47}
  (1993) 426--455, [\href{http://arxiv.org/abs/hep-ph/9207245}{{\tt
  hep-ph/9207245}}].

\bibitem{Kim:2004rp}
J.~E. Kim, H.~P. Nilles, and M.~Peloso, {\it {Completing natural inflation}},
  {\em JCAP} {\bf 0501} (2005) 005,
  [\href{http://arxiv.org/abs/hep-ph/0409138}{{\tt hep-ph/0409138}}].

\bibitem{Dimopoulos:2005ac}
S.~Dimopoulos, S.~Kachru, J.~McGreevy, and J.~G. Wacker, {\it {N-flation}},
  {\em JCAP} {\bf 0808} (2008) 003,
  [\href{http://arxiv.org/abs/hep-th/0507205}{{\tt hep-th/0507205}}].

\bibitem{Silverstein:2008sg}
E.~Silverstein and A.~Westphal, {\it {Monodromy in the CMB: Gravity Waves and
  String Inflation}},  {\em Phys.Rev.} {\bf D78} (2008) 106003,
  [\href{http://arxiv.org/abs/0803.3085}{{\tt arXiv:0803.3085}}].

\bibitem{Kaloper:2008fb}
N.~Kaloper and L.~Sorbo, {\it {A Natural Framework for Chaotic Inflation}},
  {\em Phys.Rev.Lett.} {\bf 102} (2009) 121301,
  [\href{http://arxiv.org/abs/0811.1989}{{\tt arXiv:0811.1989}}].

\bibitem{Ross:2009hg}
G.~G. Ross and G.~German, {\it {Hybrid natural inflation from non Abelian
  discrete symmetry}},  {\em Phys.Lett.} {\bf B684} (2010) 199--204,
  [\href{http://arxiv.org/abs/0902.4676}{{\tt arXiv:0902.4676}}].

\bibitem{Kaloper:2011jz}
N.~Kaloper, A.~Lawrence, and L.~Sorbo, {\it {An Ignoble Approach to Large Field
  Inflation}},  {\em JCAP} {\bf 1103} (2011) 023,
  [\href{http://arxiv.org/abs/1101.0026}{{\tt arXiv:1101.0026}}].

\bibitem{Pajer:2013fsa}
E.~Pajer and M.~Peloso, {\it {A review of Axion Inflation in the era of
  Planck}},  {\em Class.Quant.Grav.} {\bf 30} (2013) 214002,
  [\href{http://arxiv.org/abs/1305.3557}{{\tt arXiv:1305.3557}}].

\bibitem{Kehagias:2014wza}
A.~Kehagias and A.~Riotto, {\it {Remarks about the Tensor Mode Detection by the
  BICEP2 Collaboration and the Super-Planckian Excursions of the Inflaton
  Field}},  {\em Phys.Rev.} {\bf D89} (2014) 101301,
  [\href{http://arxiv.org/abs/1403.4811}{{\tt arXiv:1403.4811}}].

\bibitem{Anber:2006xt}
M.~M. Anber and L.~Sorbo, {\it {N-flationary magnetic fields}},  {\em JCAP}
  {\bf 0610} (2006) 018, [\href{http://arxiv.org/abs/astro-ph/0606534}{{\tt
  astro-ph/0606534}}].

\bibitem{Anber:2009ua}
M.~M. Anber and L.~Sorbo, {\it {Naturally inflating on steep potentials through
  electromagnetic dissipation}},  {\em Phys.Rev.} {\bf D81} (2010) 043534,
  [\href{http://arxiv.org/abs/0908.4089}{{\tt arXiv:0908.4089}}].

\bibitem{Barnaby:2010vf}
N.~Barnaby and M.~Peloso, {\it {Large Nongaussianity in Axion Inflation}},
  {\em Phys.Rev.Lett.} {\bf 106} (2011) 181301,
  [\href{http://arxiv.org/abs/1011.1500}{{\tt arXiv:1011.1500}}].

\bibitem{Nurmi:2013gpa}
S.~Nurmi and M.~S. Sloth, {\it {Constraints on Gauge Field Production during
  Inflation}},  \href{http://arxiv.org/abs/1312.4946}{{\tt arXiv:1312.4946}}.

\bibitem{Hannestad:2009yx}
S.~Hannestad, T.~Haugbolle, P.~R. Jarnhus, and M.~S. Sloth, {\it
  {Non-Gaussianity from Axion Monodromy Inflation}},  {\em JCAP} {\bf 1006}
  (2010) 001, [\href{http://arxiv.org/abs/0912.3527}{{\tt arXiv:0912.3527}}].

\bibitem{Flauger:2010ja}
R.~Flauger and E.~Pajer, {\it {Resonant Non-Gaussianity}},  {\em JCAP} {\bf
  1101} (2011) 017, [\href{http://arxiv.org/abs/1002.0833}{{\tt
  arXiv:1002.0833}}].

\bibitem{Sorbo:2011rz}
L.~Sorbo, {\it {Parity violation in the Cosmic Microwave Background from a
  pseudoscalar inflaton}},  {\em JCAP} {\bf 1106} (2011) 003,
  [\href{http://arxiv.org/abs/1101.1525}{{\tt arXiv:1101.1525}}].

\bibitem{Arvanitaki:2009fg}
A.~Arvanitaki, S.~Dimopoulos, S.~Dubovsky, N.~Kaloper, and J.~March-Russell,
  {\it {String Axiverse}},  {\em Phys.Rev.} {\bf D81} (2010) 123530,
  [\href{http://arxiv.org/abs/0905.4720}{{\tt arXiv:0905.4720}}].

\bibitem{Mukohyama:2014gba}
S.~Mukohyama, R.~Namba, M.~Peloso, and G.~Shiu, {\it {Blue Tensor Spectrum from
  Particle Production during Inflation}},
  \href{http://arxiv.org/abs/1405.0346}{{\tt arXiv:1405.0346}}.

\bibitem{Senatore:2011sp}
L.~Senatore, E.~Silverstein, and M.~Zaldarriaga, {\it {New Sources of
  Gravitational Waves during Inflation}},
  \href{http://arxiv.org/abs/1109.0542}{{\tt arXiv:1109.0542}}.

\bibitem{Biagetti:2013kwa}
M.~Biagetti, M.~Fasiello, and A.~Riotto, {\it {Enhancing Inflationary Tensor
  Modes through Spectator Fields}},  {\em Phys.Rev.} {\bf D88} (2013), no.~10
  103518, [\href{http://arxiv.org/abs/1305.7241}{{\tt arXiv:1305.7241}}].

\bibitem{Cook:2013xea}
J.~L. Cook and L.~Sorbo, {\it {An inflationary model with small scalar and
  large tensor nongaussianities}},  {\em JCAP} {\bf 1311} (2013) 047,
  [\href{http://arxiv.org/abs/1307.7077}{{\tt arXiv:1307.7077}}].

\bibitem{Jain:2012ga}
R.~K. Jain and M.~S. Sloth, {\it {Consistency relation for cosmic magnetic
  fields}},  {\em Phys.Rev.} {\bf D86} (2012) 123528,
  [\href{http://arxiv.org/abs/1207.4187}{{\tt arXiv:1207.4187}}].

\bibitem{Bruni:1996im}
M.~Bruni, S.~Matarrese, S.~Mollerach, and S.~Sonego, {\it {Perturbations of
  space-time: Gauge transformations and gauge invariance at second order and
  beyond}},  {\em Class.Quant.Grav.} {\bf 14} (1997) 2585--2606,
  [\href{http://arxiv.org/abs/gr-qc/9609040}{{\tt gr-qc/9609040}}].

\bibitem{Jarnhus:2007ia}
P.~R. Jarnhus and M.~S. Sloth, {\it {de Sitter limit of inflation and nonlinear
  perturbation theory}},  {\em JCAP} {\bf 0802} (2008) 013,
  [\href{http://arxiv.org/abs/0709.2708}{{\tt arXiv:0709.2708}}].

\bibitem{Barnaby:2011vw}
N.~Barnaby, R.~Namba, and M.~Peloso, {\it {Phenomenology of a Pseudo-Scalar
  Inflaton: Naturally Large Nongaussianity}},  {\em JCAP} {\bf 1104} (2011)
  009, [\href{http://arxiv.org/abs/1102.4333}{{\tt arXiv:1102.4333}}].

\bibitem{Chaicherdsakul:2006ui}
K.~Chaicherdsakul, {\it {Quantum Cosmological Correlations in an Inflating
  Universe: Can fermion and gauge fields loops give a scale free spectrum?}},
  {\em Phys.Rev.} {\bf D75} (2007) 063522,
  [\href{http://arxiv.org/abs/hep-th/0611352}{{\tt hep-th/0611352}}].

\bibitem{Jain:2012vm}
R.~K. Jain and M.~S. Sloth, {\it {On the non-Gaussian correlation of the
  primordial curvature perturbation with vector fields}},  {\em JCAP} {\bf
  1302} (2013) 003, [\href{http://arxiv.org/abs/1210.3461}{{\tt
  arXiv:1210.3461}}].

\bibitem{Gordon:2000hv}
C.~Gordon, D.~Wands, B.~A. Bassett, and R.~Maartens, {\it {Adiabatic and
  entropy perturbations from inflation}},  {\em Phys.Rev.} {\bf D63} (2001)
  023506, [\href{http://arxiv.org/abs/astro-ph/0009131}{{\tt
  astro-ph/0009131}}].

\bibitem{Barnaby:2012xt}
N.~Barnaby, J.~Moxon, R.~Namba, M.~Peloso, G.~Shiu, {\em et~al.}, {\it {Gravity
  waves and non-Gaussian features from particle production in a sector
  gravitationally coupled to the inflaton}},  {\em Phys.Rev.} {\bf D86} (2012)
  103508, [\href{http://arxiv.org/abs/1206.6117}{{\tt arXiv:1206.6117}}].

\bibitem{Ade:2013ydc}
{\bf Planck Collaboration} Collaboration, P.~Ade {\em et~al.}, {\it {Planck
  2013 Results. XXIV. Constraints on primordial non-Gaussianity}},
  \href{http://arxiv.org/abs/1303.5084}{{\tt arXiv:1303.5084}}.

\bibitem{Linde:2005he}
A.~D. Linde, V.~Mukhanov, and M.~Sasaki, {\it {Post-inflationary behavior of
  adiabatic perturbations and tensor-to-scalar ratio}},  {\em JCAP} {\bf 0510}
  (2005) 002, [\href{http://arxiv.org/abs/astro-ph/0509015}{{\tt
  astro-ph/0509015}}].

\bibitem{Sasaki:1986hm}
M.~Sasaki, {\it {Large Scale Quantum Fluctuations in the Inflationary
  Universe}},  {\em Prog.Theor.Phys.} {\bf 76} (1986) 1036.

\bibitem{Mukhanov:1988jd}
V.~F. Mukhanov, {\it {Quantum Theory of Gauge Invariant Cosmological
  Perturbations}},  {\em Sov.Phys.JETP} {\bf 67} (1988) 1297--1302.

\bibitem{Byrnes:2006fr}
C.~T. Byrnes and D.~Wands, {\it {Curvature and isocurvature perturbations from
  two-field inflation in a slow-roll expansion}},  {\em Phys.Rev.} {\bf D74}
  (2006) 043529, [\href{http://arxiv.org/abs/astro-ph/0605679}{{\tt
  astro-ph/0605679}}].

\bibitem{Meerburg:2012id}
P.~D. Meerburg and E.~Pajer, {\it {Observational Constraints on Gauge Field
  Production in Axion Inflation}},  {\em JCAP} {\bf 1302} (2013) 017,
  [\href{http://arxiv.org/abs/1203.6076}{{\tt arXiv:1203.6076}}].

\bibitem{Linde:2012bt}
A.~Linde, S.~Mooij, and E.~Pajer, {\it {Gauge field production in SUGRA
  inflation: local non-Gaussianity and primordial black holes}},  {\em
  Phys.Rev.} {\bf D87} (2013) 103506,
  [\href{http://arxiv.org/abs/1212.1693}{{\tt arXiv:1212.1693}}].

\bibitem{Enqvist:2001zp}
K.~Enqvist and M.~S. Sloth, {\it {Adiabatic CMB perturbations in pre - big bang
  string cosmology}},  {\em Nucl.Phys.} {\bf B626} (2002) 395--409,
  [\href{http://arxiv.org/abs/hep-ph/0109214}{{\tt hep-ph/0109214}}].

\bibitem{Lyth:2001nq}
D.~H. Lyth and D.~Wands, {\it {Generating the curvature perturbation without an
  inflaton}},  {\em Phys.Lett.} {\bf B524} (2002) 5--14,
  [\href{http://arxiv.org/abs/hep-ph/0110002}{{\tt hep-ph/0110002}}].

\bibitem{Moroi:2001ct}
T.~Moroi and T.~Takahashi, {\it {Effects of cosmological moduli fields on
  cosmic microwave background}},  {\em Phys.Lett.} {\bf B522} (2001) 215--221,
  [\href{http://arxiv.org/abs/hep-ph/0110096}{{\tt hep-ph/0110096}}].

\bibitem{Mollerach:1989hu}
S.~Mollerach, {\it {Isocurvature Baryon Perturbations and Inflation}},  {\em
  Phys.Rev.} {\bf D42} (1990) 313--325.

\bibitem{Linde:1996gt}
A.~D. Linde and V.~F. Mukhanov, {\it {Nongaussian isocurvature perturbations
  from inflation}},  {\em Phys.Rev.} {\bf D56} (1997) 535--539,
  [\href{http://arxiv.org/abs/astro-ph/9610219}{{\tt astro-ph/9610219}}].

\bibitem{Ade:2014xna}
{\bf BICEP2 Collaboration} Collaboration, P.~Ade {\em et~al.}, {\it {Detection
  of B-Mode Polarization at Degree Angular Scales by BICEP2}},  {\em
  Phys.Rev.Lett.} {\bf 112} (2014) 241101,
  [\href{http://arxiv.org/abs/1403.3985}{{\tt arXiv:1403.3985}}].

\bibitem{Caprini:2014mja}
C.~Caprini and L.~Sorbo, {\it {Adding helicity to inflationary
  magnetogenesis}},  \href{http://arxiv.org/abs/1407.2809}{{\tt
  arXiv:1407.2809}}.

\bibitem{Shiraishi:2013kxa}
M.~Shiraishi, A.~Ricciardone, and S.~Saga, {\it {Parity violation in the CMB
  bispectrum by a rolling pseudoscalar}},  {\em JCAP} {\bf 1311} (2013) 051,
  [\href{http://arxiv.org/abs/1308.6769}{{\tt arXiv:1308.6769}}].

\bibitem{Kenton:2014gma}
Z.~Kenton and S.~Thomas, {\it {D-brane Potentials in the Warped Resolved
  Conifold and Natural Inflation}},  \href{http://arxiv.org/abs/1409.1221}{{\tt
  arXiv:1409.1221}}.

\bibitem{Harigaya:2014ola}
K.~Harigaya, M.~Ibe, and T.~T. Yanagida, {\it {R-symmetric Axion/Natural
  Inflation in Supergravity via Deformed Moduli Dynamics}},
  \href{http://arxiv.org/abs/1409.0330}{{\tt arXiv:1409.0330}}.

\bibitem{Abe:2014pwa}
H.~Abe, T.~Kobayashi, and H.~Otsuka, {\it {Towards natural inflation from
  weakly coupled heterotic string theory}},
  \href{http://arxiv.org/abs/1409.8436}{{\tt arXiv:1409.8436}}.

\bibitem{Higaki:2014mwa}
T.~Higaki and F.~Takahashi, {\it {Axion Landscape and Natural Inflation}},
  \href{http://arxiv.org/abs/1409.8409}{{\tt arXiv:1409.8409}}.

\bibitem{Ben-Dayan:2014lca}
I.~Ben-Dayan, F.~G. Pedro, and A.~Westphal, {\it {Towards Natural Inflation in
  String Theory}},  \href{http://arxiv.org/abs/1407.2562}{{\tt
  arXiv:1407.2562}}.

\bibitem{Long:2014dta}
C.~Long, L.~McAllister, and P.~McGuirk, {\it {Aligned Natural Inflation in
  String Theory}},  {\em Phys.Rev.} {\bf D90} (2014) 023501,
  [\href{http://arxiv.org/abs/1404.7852}{{\tt arXiv:1404.7852}}].

\bibitem{Kappl:2014lra}
R.~Kappl, S.~Krippendorf, and H.~P. Nilles, {\it {Aligned Natural Inflation:
  Monodromies of two Axions}},  {\em Phys.Lett.} {\bf B737} (2014) 124--128,
  [\href{http://arxiv.org/abs/1404.7127}{{\tt arXiv:1404.7127}}].

\bibitem{Takahashi:2013tj}
F.~Takahashi, {\it {The Spectral Index and its Running in Axionic Curvaton}},
  {\em JCAP} {\bf 1306} (2013) 013, [\href{http://arxiv.org/abs/1301.2834}{{\tt
  arXiv:1301.2834}}].

\bibitem{Sloth:2014sga}
M.~S. Sloth, {\it {Chaotic inflation with curvaton induced running}},  {\em
  Phys.Rev.} {\bf D90} (2014) 063511,
  [\href{http://arxiv.org/abs/1403.8051}{{\tt arXiv:1403.8051}}].

\end{thebibliography}\endgroup

\end{document}